\documentclass[prl,twocolumn,aps,amsmath,amssymb,showpacs,superscriptaddress]{revtex4}

\usepackage{graphics}
\usepackage{dcolumn}
\usepackage{bm}
\usepackage[T1]{fontenc}

\begin{document}

\title{Shot Noise Suppression and Hopping Conduction in Graphene Nanoribbons}

\author{R. Danneau}\email[Corresponding author: romain.danneau@kit.edu]{}
\affiliation{Low Temperature Laboratory, Aalto University, Finland}
\affiliation{Institute of Nanotechnology and Institute of Physics,  Karlsruhe Institute of Technology, Germany}
\author{F. Wu}\email[Present address: Microtechnology and Nanoscience MC2, Chalmers Univ. of Techn., 41296 Göteborg, Sweden.]{}
\author{M.Y. Tomi}
\affiliation{Low Temperature Laboratory, Aalto University, Finland}
\author{J.B. Oostinga}
\affiliation{Condensed Matter Physics Department, University of Geneva, Switzerland}
\affiliation{Kavli Institute of Nanoscience, Delft University of Technology, The Netherlands}
\author{A.F. Morpurgo}
\affiliation{Condensed Matter Physics Department, University of Geneva, Switzerland}
\author{P.J. Hakonen}
\affiliation{Low Temperature Laboratory, Aalto University, Finland}

\begin{abstract}

We have investigated shot noise and conduction of graphene field effect nanoribbon devices at low temperature.
By analyzing the exponential $I-V$ characteristics of our devices in the transport gap region, we found out that transport follows
variable range hopping laws at intermediate bias voltages $1 < V_{bias} < 12$ mV. In parallel, we observe a strong shot noise suppression
leading to very low Fano factors. The strong suppression of shot noise is consistent with inelastic hopping, in crossover from one- to two-dimensional regime,
indicating that the localization length $l_{loc} < W$ in our nanoribbons.

\end{abstract}

\pacs{72.80.Vp, 73.50.Td}

\maketitle

Graphene, a two-dimensional crystal of carbon atoms, has shown some amazing electrical properties  \cite{geim2007} attracting the interest of both scientific community and microelectronic industry. However, graphene is a zero-gap semiconductor with a minimum conductivity way too large to be utilized as base material for high on-off ratio field effect transistor. One way to circumvent this problem would be to open a gap in graphene's band-structure. It is possible in bilayer graphene by the means of doping (either chemical \cite{otah2006} or electrostatic \cite{oostinga2008}).
Another way to create an efficient graphene transistor is to build a constriction and/or to form a nanoribbon.
Early theoretical studies  have predicted that a gap could be opened in graphene nanoribbons (GNR) depending on the edges being either
zigzag or armchair \cite{nakada1996}. 

However, the first studies of GNRs were performed on etched graphene leading to ribbon width down to around 20 nm \cite{chen2007,han2007}. These experiments demonstrated the presence of a transport gap inversely proportional to the width and independent on the crystallographic orientation \cite{han2007}. It was also estimated that part of the ribbons at the edges were probably not conducting (around 14 nm at $T$ = 4.2 K), suggesting that edge roughness is significant. Similar transport gaps were observed for much smaller ribbon width in GNRs fabricated using sonication of intercalated graphite in solution, indicating smoother edges than the etched GNRs \cite{li2008}. Indeed, experiments performed on GNRs \cite{chen2007,han2007,ozyilmaz2007,han2009,oostinga2010} tend to prove that the origin of the gap may be more complex than the early theoretical studies suggested \cite{nakada1996}. Despite several models based on Anderson localization, Coulomb blockade or percolation phenomenon \cite{son2006}, there is not yet a consensus as to the origin of the gap in GNRs.

In this work, we report the first shot noise measurements on etched GNRs performed at low temperature.
Our results show a strong shot noise reduction while $I-V$ characteristics measured follow variable range hopping (VRH) laws \cite{shklovskii1984} in the gap region.
Such shot noise suppression is the consequence of inelastic hopping conduction from a localized state to an adjacent one, localized states arising from the rough
edges and disorder due to residues and defects from the fabrication process. We also find that relaxation of electrons is stronger than expected in our ribbons.

The GNRs have been fabricated from the same graphene monolayer (identified using the RGB green shift as described in \cite{oostinga2008,craciun2009,oostinga2010}) using Scotch tape micromechanical cleavage on natural graphite. The graphene sheets were deposited on a heavily \emph{p}-doped substrate with 300 nm SiO$_2$ layer (see Fig. 1(a)).
The graphene sheet was first connected using standard e-beam lithography followed by a Ti(10 nm)/Au(40 nm) bilayer deposition with lift-off in acetone.
A second lithography step allowed the patterning of the GNRs. The resist (PMMA) was used as mask in this step and GNRs were etched using an Ar plasma. We present the measurements on two GNRs: Sample A with a length $L \sim 600$ nm and a nominal width $W \sim 90$ nm, and sample B with a length $L \sim 200$ nm and a nominal width $W \sim 70$ nm. After the experiments, the GNRs were observed using scanning electron microscope at 0.5 kV (see Fig. 1(b)).

\begin{figure}[htbp]
\scalebox{0.3}{\includegraphics{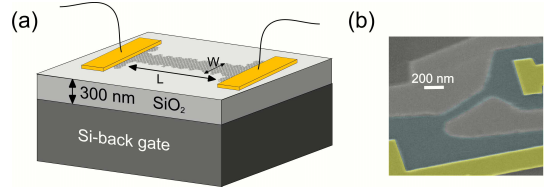}} {\caption{(a) Schematics of an etched GNR. (b) False color scanning electron micrograph of sample A, highlighting the graphene (in blue) and the Ti/Au contacts (in yellow).}}
\end{figure}

The measurements were performed in a similar fashion as described in Ref. \onlinecite{danneau2008}, from room temperature down to $T$ = 4.2 K. The differential conductance $\frac{dI}{dV}$ was measured using standard low-frequency ac lock-in technique with an excitation amplitude from 0.38 mV up to 0.8 mV ($\sim$ 4 K to $\sim$ 8 K) at \emph{f}= 63.5 Hz. A tunnel junction was used for calibration of the shot noise \cite{danneau2008,wu2006}.

Fig. 2(a) and (b) display the gate voltage $V_{gate}$ dependence of the zero bias conductance $G$ for different temperatures $T$ of sample A and B, respectively.
In both cases, we observe a drop of $G$ when $T$ is lowered, and a high impedance region emerges as $T \rightarrow 4.2$ K.
Clear conductance oscillations at zero bias are visible at the lowest temperatures. However, no periodicity is detectable in a Fourier analysis.
Far away from the charge neutrality point $G \sim 2e^2/h$, i.e. twice the conductance quantum $g_0$.
On Fig 2(c) and (d), we show a color map of the scaled differential conductance $\frac{dI}{dV}/g_0$ as a function of bias voltage $V_{bias}$ and $V_{gate}$ at liquid helium temperature, for sample A and B, respectively. These measurements highlight the formation of a "large impedance region" or a "gap" as previously observed \cite{chen2007,han2007,ozyilmaz2007,han2009,oostinga2010}. This region can be viewed in different ways. In the Anderson picture, it arises from localization due to the rough edges and the disorder resulting in to the high impedance region (around the original Dirac point) at zero bias.
Out of equilibrium measurements, on the other hand, illuminate the Coulombic aspects of the transport suppression in GNRs: a "source and drain" gap is modulated by the "Coulomb diamond-like" structures which could originate from the formation of a series of dots, all contributing their share to the "gap". We found "source drain gap" of about 5 meV and 15 meV from our $\frac{dI}{dV}-V_{bias}$ data, and a "transport gap" of about 14 V and 18 V from the $\frac{dI}{dV}-V_{gate}$ curves for sample A and B respectively. We observe clear irregular "Coulomb diamond"-like structures comparable to previous studies \cite{chen2007,han2007,ozyilmaz2007,han2009,oostinga2010}, suggesting that Coulomb interactions are significant.

\begin{figure}[htbp]
\scalebox{0.43}{\includegraphics{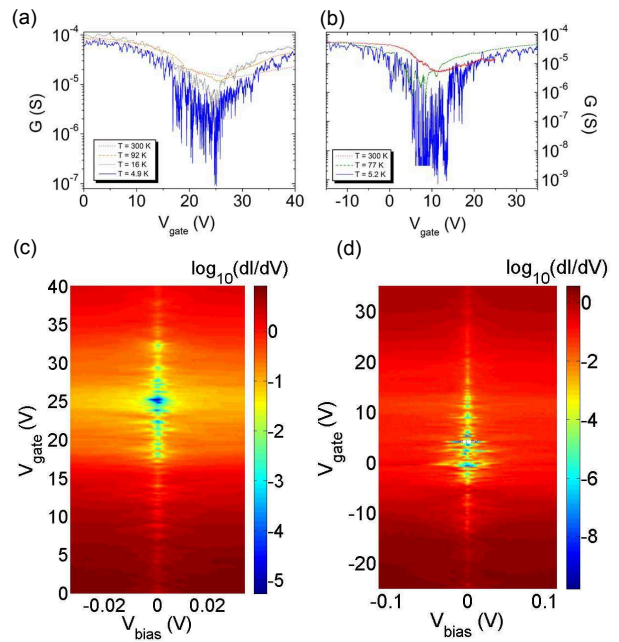}} {\caption{(a) and (b): $G$ versus $V_{gate}$ at various temperature for sample A and B respectively. (c) and (d): color map of t $\frac{dI}{dV}$  versus $V_{bias}$ and $V_{gate}$ at $T =$ 4.9 K for sample A and 5.2 K for sample B respectively.}}
\end{figure}

VRH generally describes electronic transport in the presence of disorder \cite{shklovskii1984}. Temperature dependence of the conductance $G(T)$ is conventionally used to identify the regime. In the case of GNRs, the minimum conductance can vary in gate voltage $V_{gate}$ as the temperature is lowered even under vacuum condition \cite {han2007}, leading to uncertainties in the data analysis. The uncontrolled doping by adsorbed molecules may move the minimum conduction region during the cool down. However, G(T) study has been recently successfully performed \cite{han2009}. An alternative way is to analyze $I-V$ curves at a temperature $T$. At high bias, below a certain $V_0$, the following equation can be used to describe VRH:
\begin{eqnarray}
I(E,T) = VG_{0}(T) \exp \left\{-\left(\frac{V_0}{V} \right)^{1/(d+1)} \right\} \label{equ1}
\end{eqnarray}
where $d$ is the dimensionality of hopping (for the effect of interactions, see below) and $G_{0}$ is the zero bias conductance. Eq. \ref{equ1} transforms to Mott's law by replacement of $e V_0= k_B T_0$ and $e V = k_B T$ in the exponent ($V_0$ being the upper most value for which the formula is valid) which provides the basic motivation for using this functional form \cite{pollak,likharev,fogler}.

\begin{figure}[htbp]
\scalebox{0.29}{\includegraphics{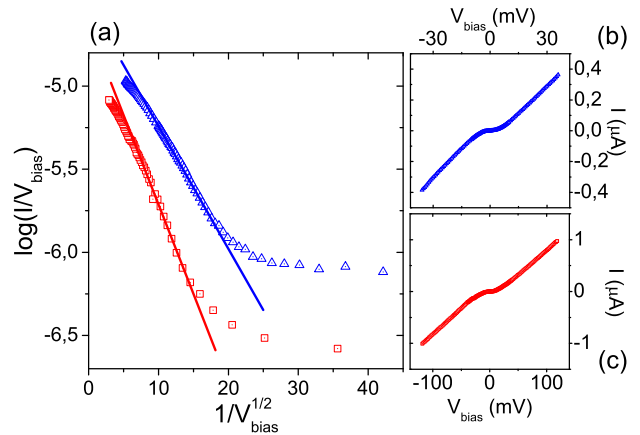}} {\caption{\emph{I-V} characteristics of sample A and B plotted using hopping law at high bias (a), eq. (\ref{equ1}), at $V_{gate}$ = 25.4 and 11 V at $T$ = 4.9 and 5.2 K for sample A and B respectively ($\vartriangle$ and $\square$). The plot shows linear behavior in the log-scale with $1/V_{bias}^{1/2}$ above the gap (flat part of (b) and (c)) and below $V_0$, i.e. where the data stats to deviate from linear in (a). (b) and (c) are the corresponding normal \emph{I-V} plots of sample A and B respectively.}}
\end{figure}

Fig. 3 displays $I-V$ curves for sample A and B measured in the gap region. Following Eq. \ref{equ1}, we see that the conduction follows variable range hopping law in the gap region.
The data are plotted using $d=1$ which describes VRH for one-dimensional (1-d) systems with or without interactions or two-dimensional (2-d) systems with interactions.
We obtain $V_0 \sim$ 8 and 12 mV for sample A and B, respectively.
Here, $\frac{a}{L}e V_0$ describes the bias needed to overcome the potential barrier of the localized state with radius $a$.
The fact that we obtain a larger $V_0$ for sample B which has a width 20 nm smaller (and is even shorter) than sample A indicates an enhanced influence of the rough edges on the conduction. Consequently, our results show that the appearance of the high impedance region in GNRs is also affected by defects like localized states at the edges and, likewise, by the local doping due to contaminants. This is in agreement with the recent works on temperature dependence of GNR conductance \cite{han2009}. Han \emph{et al.} have shown that for various GNR geometries  $l \gtrsim W$ indicating 1-d VRH transport in the high impedance region of GNRs; the origin of the transport gap would then be due to localized states \cite{han2009}. This has recently been confirmed by magneto-transport measurements \cite{oostinga2010}. Our value for $V_0 \simeq 10$ meV is close to the value $k_B T_0/e \simeq 6$ meV given in Ref. \onlinecite{oostinga2010}.

In order to gain more information on the hopping in GNRs, we have studied shot noise. Shot noise denotes current fluctuations arising from the granular nature of the charge carriers (see Ref. \onlinecite{blanter2000} for a review). It provides a powerful tool to probe mesoscopic systems and
it is usually regarded as a complementary technique to conductance measurements. The Fano factor $F$, given by the ratio of shot noise and mean current, is commonly
employed to quantify shot noise. The noise power spectrum then reads $S(I) = F \times 2eI$.
In the case of phase coherent transport in GNRs, shot noise strongly depends on the boundary conditions, i.e. whether the edges are zigzag or armchair \cite{cresti2007}.
However, phase coherent length in etched GNRs have been estimated to be at most 175 nm \cite{oostinga2010} and it is clearly less in our experiment due to higher temperature and a finite bias that enhances energy relaxation. While in the  case of phase coherent transport, shot noise can be described simply by the scattering matrix theory, it can be treated using semiclassical means in the incoherent regime. When inelastic processes dominate (inelastic length $l_{in} <L$), shot noise starts to decrease and it becomes dependent on the details of the relaxation processes that govern the ensuing non-equilibrium state. In inelastic hopping conduction with short hopping length ($l_{hop}<<L$), strong suppression of shot noise takes place as observed in \cite{kuznetsov2000}.

Assuming strongly inelastic behavior, classical addition of uncorrelated noise sources can be employed and networks of resistors with shunting current noise generators become an appealing choice for noise modeling in GNRs. Within this classical limit, the internal topology of the ribbon becomes relevant. If hopping is 2-d in GNRs, then part of the noise current of individual noise generators is shunted via the conduction paths inside the ribbon and the noise coupled to an outside load becomes reduced. Consequently, we expect that the Fano factor is reduced a bit further down from the 1-d classical limit given by $l_{hop}/L$.

We have performed our shot noise measurements at frequency around 800 MHz. This frequency is high enough so that all noise due to slow fluctuations of resistance (transmission coefficients) can be neglected. On the other hand, the frequency is low compared with internal charge relaxation time scales and high frequency effects can be neglected. Fig. 4 displays the current noise per unit bandwidth $S_I$ versus current $I$ in the high impedance
region for samples A and B, respectively. Both curves are fitted using the formula defined previously \cite{danneau2008} with $F$ as the only fitting parameter.
We find a rather low Fano factor for both GNRs $F \sim 0.1$ at low bias (the results involve a correction due to non-linear $I-V$ curves as discussed in Ref. \onlinecite{danneau2008}).
With increasing bias, we find a further reduction of the Fano factor, which signals a strong role of inelastic processes as the localized states become delocalized.

\begin{figure}[htbp]
\scalebox{0.29}{\includegraphics{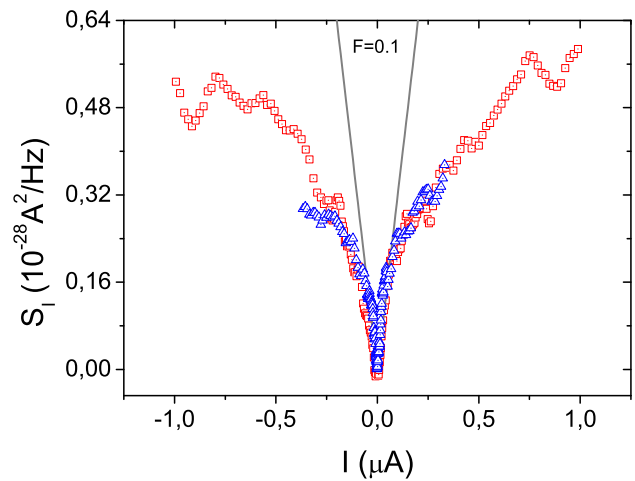}} {\caption{$S_I$ versus $I$ averaged over three gate values around at $V_{gate}$ = 25.4 V
and $T$ = 4.9 K for sample A ($\vartriangle$), and $V_{gate}$ = 11 V and $T$ = 5.2 K for sample B ($\square$). We show a low-bias fit for sample B using the Khlus formula with $F$ as the only fitting parameter. $F$ decreases at higher bias (see text).}}
\end{figure}

Why such a low shot noise?
The observed conductance modulation in the high impedance regime suggests that a series/array of dots is formed in GNRs. Quantum dots often show super-Poissonian noise instead of low noise level (see, for example, the work done on carbon nanotubes \cite{onac2006,wu2007}) and as theoretically expected for a series of quantum dots \cite{aghassi2006}. However, a series of $N$ quantum dots without inelastic effects should lead to a Fano factor of $\frac{1}{3}$  \cite{golubev2004}. We note that shot noise suppression could be seen in asymmetric, open quantum cavity \cite{blanter2000}, but the resistance of one or two open quantum cavities (regions at the ends of the ribbon) is too small to account for our results. There will, however, be a small contribution by the end reservoirs on the shot noise.

The main contribution to the shot noise suppression can only come from hopping conduction via so small localized states that the nature of hopping conduction is likely to be almost 2-d. $F$ for a series of $N$ sites with inelastic hopping is approximately
$1/N \sim l_{hop}/L$, and this remains as a good approximation also in the 2-d situation where N then denotes the number of hops \emph{along the voltage bias}.
In order to explain the observed suppression,
the hopping length has to be in the range of $l_{hop} \sim 20-60$ nm; As the localization length $l_{loc} \sim l_{hop}$ is less than the width of the GNR, we conclude that the hopping conduction in our ribbons
is not 1-d in nature but rather it falls in the crossover regime between 1-d and 2-d (or quasi 1-d).
Our shot noise results thus indicate even a slightly smaller hopping length than was found in \cite{han2009,oostinga2010}.

The shot noise crossover from VRH region to high bias regime without localized states in Fig. 4 points to strong relaxation of electrons: otherwise an increase of the Fano factor would be expected across
the crossover as the number of hops decreases and $l_{loc}$ increases \cite{kuznetsov2000,likharev}. Indeed, even in the VRH regime, the apparent Fano factor could be formed by other means,
for example by noise from the graphene islands at the ends, and that the actual shot noise from the ribbon nearly vanishes. This would be reminiscent to carbon nanotubes where very small $F$ have been observed in various configurations \cite{roche2002,tsuneta2009}. Nearly total suppression of shot noise indicates very effective energy relaxation at finite bias which could be realized by disorder-enhanced electron-phonon coupling \cite{sergeev2000} or by relaxation via new degrees of freedom provided by the edges of the ribbon.

To conclude, we have measured shot noise and conductance in GNRs. While the dc transport shows characteristic behavior of GNRs, we clearly observe a strong shot noise suppression. We were able to fit the $I-V$ curves with VRH laws in the high impedance region. We have shown that shot noise suppression could be explained by inelastic hopping conduction in the quasi-1-d limit. Our results are consistent with the strong effect of rough edges and local contaminants in the conduction and shot noise of GNRs.
Shot noise being a limited factor for electrical devices, our findings enlighten the great potential of GNRs as building blocks for future electronics.

\begin{acknowledgments}
We thank F. Evers, D. Golubev, T. Heikkilä, M. Hettler, A. Ioselevich, P. Pasanen, M. Laakso, G. Metalidis and M. Paalanen for fruitful discussions. This work was funded by the Academy of Finland, by EU Commission (FP6-IST-021285-2) and by Nokia NRC (NANOSYSTEMS project). R.D.'s Shared Research Group SRG 1-33 received financial support by the "Concept for the future" of the Karlsruhe Institute of Technology within the framework of the German Excellence Initiative.
\end{acknowledgments}

\end{document}